\documentclass[epj]{webofc}
\usepackage[varg]{txfonts}   
\wocname{EPJ Web of Conferences}
\woctitle{CONF12}
\begin{document}
\selectlanguage{english}
\title{Simplified Limits on New LHC Resonances}

\author{Elizabeth H. Simmons\inst{1}\fnsep\thanks{\email{esimmons@msu.edu}} \and
        R. Sekhar Chivukula\inst{1} \and
        Pawin Ittisamai\inst{2}\and 
        Kirtimaan Mohan\inst{1}
}

\institute{Department of Physics and Astronomy, 
Michigan State University, East Lansing U.S.A. 
\and
Department of Physics, Faculty of Science, 
Chulalongkorn University, Bangkok 10330, Thailand
}

\abstract{%
If an excess potentially heralding new physics is noticed in collider data, it would be useful to be able to compare the data with entire classes of models at once.  This talk discusses a method that applies when the new physics corresponds to the production and decay of a single, relatively narrow, $s$-channel resonance. A simplifed model of the resonance allows us to convert an estimated signal cross section into model-independent bounds on the product of the branching ratios corresponding to production and decay. This quickly reveals whether a given class of models could possibly produce a signal of the observed size.  We will describe how to apply our analysis framework to cases of current experimental interest, including resonances decaying to dibosons, diphotons, dileptons, or dijets.  
}
\maketitle
\section{Introduction}
\label{sec:intro}

LHC data is being used to seek evidence of $2\to 2$ scattering processes where a narrow resonance arising from physics Beyond the Standard Model (BSM) is produced in the $s$-channel and immediately decays to visible final state particles.  Experiment is often compared with theory by showing how a few benchmark models with specific parameter choices compare to the observed limits on the cross-section ($\sigma$) times branching fraction ($BR$) for the process. Yet at the beginning of the analysis, especially when a small excess may be present, it would be invaluable to compare the data with entire classes of models, to immediately pare down the list of relevant options. This work builds off of our previous results on identifying the color \cite{Atre:2013mja,Chivukula:2014npa} and spin \cite{Chivukula:2014pma} properties of new discovered resonances decaying to dijet final states.   We have extended these ideas to a wider variety of final states and to situations in which just a small deviation possibly indicative of a resonance has been observed.

Most ``model-independent" analyses reported in searches for narrow BSM resonances are now cast as a plot of the experimental upper limit on $\sigma \cdot BR$ plotted as a function of the mass of the new resonance. Theoretical prediction curves are overlaid on the data, each corresponding to a different choice of spin, electric charge, weak charge, and color charge for the new resonance; in that very general sense, the set of curves might be thought to span the theoretical possibilities.  But for a given choice of spin and charges, there would actually be many possible theoretical realizations corresponding to very different strengths and chiralities of the resonance's couplings to the partons through which it is produced and to the final states into which it decays. 

In \cite{Chivukula:2016hvp} we showed that reporting collider searches for BSM resonances in terms of a new variable $\zeta$ would quickly reveal whether an entire class of resonances with particular production modes and/or decay patterns (e.g., a spin-zero state produced through gluon fusion and decaying to diphotons) could conceivably be responsible for a given deviation in cross-section data relative to standard model predictions.  When the answer is ``no", one can move on immediately. When the answer is ``yes," one also obtains information on the range of masses and branching fractions a model must yield to be compatable with the data; this could guide model-building into profitable directions.

Evidence for or observation of an excess would generally be reported within a specific channel or set of a few channels.  Often, the BSM ideas invoked to explain the excess correspond to the production and decay of a single, relatively narrow, s-channel resonance. A simplifed resonance model lets us convert any estimated signal cross section into general constraints on the properties of the resonance. If production occurs dominantly through a single process, we can obtain model-independent upper bounds on the product of the branching ratios corresponding to production and decay for that process. This reveals whether a given class of models could possibly produce a signal of the required size at the LHC.  One can readily extend this to situations with more than one production or decay channel.

If the LHC experiments were to report their searches for new resonances beyond the standard model in the simplified limits variable $\zeta$ defined in \cite{Chivukula:2016hvp} and summarized here, it could be far easier to avoid blind alleys and home in on the most likely candidate models to explain any observed excesses.
 
\section{Narrow Resonances}
\label{sec:narrow}

As noted in \cite{Chivukula:2016hvp}, the tree-level partonic production cross-section for an arbitrary 
$s$-channel resonance $R$ produced by collisions of particular initial state partons $i,j$ and decaying to  a single final state $x,y$ at the LHC can be written \cite{Harris:2011bh,Agashe:2014kda}
\begin{equation}
\hat{\sigma}_{ij\to R\to xy}(\hat{s}) = 16 \pi (1 + \delta_{ij}) \cdot {\cal N} \cdot
\frac{\Gamma(R\to i+j) \cdot \Gamma(R\to x+y)}
{(\hat{s}-m^2_R)^2 + m^2_R \Gamma^2_R} ~,
\end{equation}
where ${\cal N}$ is a ratio of spin and color counting factors\footnote{We note here that ${\cal N}$ depends on the color and spin properties of the incoming partons $i,j$. We will neglect this in what follows, assuming that this factor is the same for all relevant production modes in a given situation -- see discussion at the end of subsection \ref{subsec:general-case}. In fact, this assumption is valid in the great majority of cases.}
\begin{equation}
{\cal N} = \frac{N_{S_R}}{N_{S_i} N_{S_j}} \cdot
\frac{C_R}{C_i C_j},
\label{eq:N}
\end{equation}
where $N_S$ and $C$ count the number of spin- and color-states for initial state partons $i$ and $j$ and for the resonance $R$. In the narrow-width approximation, one can simplify this further, using the expression\footnote{In detail, resonance limits derived from observations will depend on whether $\Gamma_R/m_R$ lies below the experimental resolution for the invariant mass of the final state particles. As discussed in \cite{Chivukula:2016hvp}, these effects are too small to be relevant here.}  
\begin{equation}
\frac{1}
{(\hat{s}-m^2_R)^2 + m^2_R \Gamma^2_R}
\approx \frac{\pi}{m_R \Gamma_R} \delta(\hat{s} - m^2_R)~.
\end{equation}

\noindent Integrating over parton densities, and summing over incoming partons and over the outgoing partons yielding experimentally indistinguishable final states, we find the tree-level hadronic cross section
\begin{align}
\sigma^{XY}_R &\ \equiv \sigma_R \times BR(R \to X + Y) = 16\pi^2 \cdot {\cal N} \cdot \frac{ \Gamma_R}{m_R} \times  \nonumber \\
& 
\left( \sum_{ij} (1 + \delta_{ij}) BR(R\to i+j) \left[\frac{1}{s} \frac{d L^{ij}}{d\tau}\right]_{\tau = \frac{m^2_R}{s}}\right) \cdot \left(\sum_{xy\, \in\, XY} BR(R\to x+y)\right)~. 
\label{eq:cross-section}
\end{align}
Here ${d L^{ij}}/{d\tau}$ corresponds to the luminosity function for the $ij$ combination of partons\footnote{
In particular,
	\begin{equation}
	\left[ \frac{d{L}^{ij}}{d\tau}\right] \equiv 
	\frac{1}{1 + \delta_{ij}} \int_{\tau}^{1} \frac{dx}{x}
			\left[ f_i\left(x, \mu_F^2\right) f_j\left( \frac{\tau}{x}, \mu_F^2 \right) +
			f_{j}\left(x, \mu_F^2\right) f_i\left( \frac{\tau}{x}, \mu_F^2 \right) \right]  \,,
	\label{eq:lumi-fun}
	\end{equation}
	where here, for the purposes of illustration, we calculate these parton luminosities using the {\tt CTEQ6L1}~\cite{Pumplin:2002vw} parton density functions, setting the factorization scale $\mu_F^2= m^2_R$. More details are given in \cite{Chivukula:2016hvp}.}, and $X\, Y$ label the
	set of experimentally indistinguishable final states.
	
This way of writing the cross-section lends itself well to judging which classes of models are capable of producing a given observable excess.  We will now walk through a few scenarios in principle and in Section \ref{sec:applications}, we will treat specific instances of these scenarios in more detail.

While we have been calculating total cross-sections, some experimental results are given as limits on the cross-section times acceptance due to kinematic cuts. Where we have encountered the latter, we have used simulations performed with {\tt MadGraphMC@NLO}~\cite{Alwall:2014hca} to evaluate the acceptance.\footnote{The acceptance due to kinematic cuts depends on the angular distribution of the final states, which  depends on the spin of the particles involved. In cases with multiple production and decay modes, the acceptance can change depending on the spins of the initial and final states \cite{Jacob:1959at,Haber:1994pe}. If there are multiple production modes with substantially different acceptances, one would have to consider these modes seperately. this does not affect the examples considered here. }

\subsection{Simplest Case: one production and one decay mode}
\label{subsec:simplest}

We start with the simplest possible case in which only one set of initial $(i,j)$ and final $(x,y)$ states is relevant for production and decay of a new resonance $R$.  

We can write down the signal cross-section for pp-collisions as follows (here $XY$ reduces to $xy$ because there is only one decay mode),
	\begin{equation}
	\sigma^{xy}_R=\sigma_R \times BR(R \to x+y) = \int_{s_{min}}^{s_{max}}d\hat{s}\,
	\hat{\sigma}(\hat{s}) \cdot \left[ \frac{d L^{ij}}{d\hat{s}}\right]~,
	\label{eq:simplest1}
	\end{equation}
and hence, in the narrow-width approximation,
\begin{equation}
\sigma^{xy}_R = 16\pi^2 \cdot {\cal N} \cdot \frac{ \Gamma_R}{m_R}  \cdot
 (1 + \delta_{ij})BR(R\to i+j) \left[\frac{1}{s} \frac{d L^{ij}}{d\tau}\right]_{\tau = \frac{m^2_R}{s}}  \cdot BR(R\to x+y)~.
\label{eq:simplest2}
\end{equation}

This can be reframed as an expression for the product of branching ratios:
\begin{equation}
   BR(R\to i+j) (1 + \delta_{ij}) \cdot BR(R\to x+y)  =  \frac{\sigma^{xy}_R}{16 \pi^2  {\cal N} \frac{\Gamma_R}{m_R} \left[\frac{1}{s} \frac{d L^{ij}}{d\tau}\right]_{\tau = \frac{m^2_R}{s}}}~.
\label{eq:simplebound-lower}
\end{equation}
This equation essentially tells us that if an {\it arbitrary} $s$-channel resonance with a given value of $\Gamma_R/m_R$ produced from partons $ij$ is to produce a signal of a particular size, then the product of the resonance's branching ratios must attain a certain value. Significantly, this value depends only on the properties of the resonance and the partonic luminosity of the initial state partons.  It can therefore be used to distinguish among potential theoretical descriptions of any new resonance.

At the same time, since the sum of all branching ratios of a resonance equals one, we can set a theoretical upper bound on the value of the product of branching ratios discussed above. There are four possibilities, depending on whether the incoming two partons are identical to one another and whether the incoming and outgoing states are the same:

\begin{equation}
BR(R\to i+j) (1 + \delta_{ij}) \cdot BR(R\to x+y)  \le
\begin{cases}
1/4 & i\neq j,\, ij \neq xy \\
1 & i\neq j,\, ij = xy\\
1/2 & i=j,\, x=y,\, ij\neq xy \\
2 & i=j, x=y, ij = xy
\end{cases}
\label{eq:combined}
\end{equation}

Experimental searches for a narrow resonances $R\to x+y$ are generally reported in terms of expected and observed upper bounds in the $\sigma^{xy}_R \equiv \sigma(pp \to R)\cdot  BR(R\to x+y)$ vs. $m_R$ plane. A potential narrow resonance appears initially (prior to a 5$\sigma$ discovery) as a deviation in which the observed limit is weaker than the expected limit. When such a deviation is seen, one immediately asks what kinds of resonances $R \to x+y$ could potentially explain this excess.  The tendency has been to make comparisons with very specific models.  

We suggest that a more general approach based on Eqn. \ref{eq:simplebound-lower} can be far more informative. Specifically, the value of the product of branching ratios required to achieve a given $\sigma_R$ can be plotted on the same plane for various choices of $ij$ and $R$ and compared with the upper bounds on that product of branching ratios.  One will immediately see which classes of resonances could potentially give rise to the observed deviation.  We will illustrate this in detail in Section \ref{sec:applications}.

\subsection{General Case: Multiple production and decay modes }
\label{subsec:general-case}

This situation is complicated by the fact that the branching ratio for each initial state ($ij$) is associated with the luminosity function for that particular pair of partons.   We will need to rewrite Eqn.~\ref{eq:cross-section} in order to relate theoretical upper limits on products of branching ratios to the value of the cross-section, resonance properties, and parton luminosities.

The sum over branching ratios times luminosities for incoming partons $ij$ in Eqn.~\ref{eq:cross-section} may be usefully reframed by simultaneously multiplying and dividing it by $ \sum_{i'j'} (1+\delta_{i'j'}) BR(R\to i' + j')$ 
\begin{align}
\sum_{ij} & (1 + \delta_{ij}) BR(R\to i+j) \left[\frac{1}{s} \frac{d L^{ij}}{d\tau}\right]_{\tau = \frac{m^2_R}{s}} = \\
&\left[\sum_{ij} \omega_{ij} \left[\frac{1}{s} \frac{d L^{ij}}{d\tau}\right]_{\tau = \frac{m^2_R}{s}}\right] \cdot \left[\sum_{i'j'} (1 + \delta_{i'j'}) BR(R\to i'+j') \right] \nonumber 
\label{eq:rewritten}
\end{align}
where
\begin{equation}
\omega_{ij} \equiv \dfrac {(1 + \delta_{ij})  BR(R\to i+j)} {\sum_{i'j'} (1 + \delta_{i'j'}) BR(R\to i'+j')}~.
\end{equation}
The fraction $\omega_{ij}$ lies in the range $ 0 \le \omega_{ij} \le 1$ and by construction $\sum_{ij} \omega_{ij} = 1$.  
Essentially, $\omega_{ij}$ tells us the weighting of each set of parton luminosities $L^{ij}$.

Inserting this in Eqn.~\ref{eq:cross-section} and re-arranging to give an expression for the product of the sums of incoming and outgoing branching ratios, we find:
\begin{align}
\left[\sum_{i'j'} (1 + \delta_{i'j'}) BR(R\to i'+j') \right]  &\cdot \left(\sum_{xy\, \in\, XY} BR(R\to x+y)\right) = 
\label{eq:gen-bran-init}\\
&\frac{\sigma^{XY}_R} { 16\pi^2 \cdot {\cal N} \cdot \frac{\Gamma_R}{m_R} \times 
\left[\sum_{ij} \omega_{ij} \left[\frac{1}{s} \frac{d L^{ij}}{d\tau}\right]_{\tau = \frac{m^2_R}{s}}\right]} ~, \nonumber
\end{align}
which generalizes Eqn. \ref{eq:simplebound-lower}. The upper bound on the product of sums over branching ratios will be $1/4$, $1/2$, $1$ or $2$, depending on the identities of the incoming ($i'j'$) and outgoing ($x,y$) partons, in a straightforward generalization of Eqn. \ref{eq:combined}.

Note that our analysis has implicitly assumed that all relevant production modes of a given resonance have the same color and spin properties (i.e., the same value of ${\cal N}$ defined in Eq. \ref{eq:N}). As discussed in \cite{Chivukula:2016hvp}, when one takes parton luminosities into account production modes with one ${\cal N}$ value usually dominate in a given search. Only rarely would one need to undertake a more sophisticated analysis simultaneously involving two production modes with different ${\cal N}$ factors.

\subsection{Simplified Language}
\label{subsec:simplified}

It actually easier to make comparisons between data and theory if one re-arranges Eqn.~\ref{eq:gen-bran-init} (and analogously  Eqn.~\ref{eq:simplebound-lower}) slightly so that the left-hand side includes the ratio of resonance width to mass. In \cite{Chivukula:2016hvp}, we thereby defined a useful dimensionless quantity called $\zeta$:
\begin{align}
\zeta \equiv \left[\sum_{i'j'} (1 + \delta_{i'j'}) BR(R\to i'+j') \right]  &\cdot \left(\sum_{xy\, \in\, XY} BR(R\to x+y)\right) \cdot \frac{\Gamma_R}{m_R} = 
\label{eq:gen-bran-rat}
\\
&\frac{\sigma^{XY}_R} { 16\pi^2 \cdot {\cal N} \times 
\left[\sum_{ij} \omega_{ij} \left[\frac{1}{s} \frac{d L^{ij}}{d\tau}\right]_{\tau = \frac{m^2_R}{s}}\right]} ~. \nonumber
\end{align}
Because we are working in the narrow width approximation, and assuming $\Gamma/M \leq 10\%$, the upper bounds on the products of branching ratios correspond to upper limits on $\zeta$ a factor of ten smaller.  

\section{Applications}
\label{sec:applications}

We will now apply the simplified limits technique to various situations of general theoretical and experimental interest.  These will include scalars decaying to diphotons, dijets, or $t\bar{t}$; a spin-1 state decaying to dibosons or dijets; a spin-2 state decaying to diphotons or dijets; a $W^\prime$ boson decaying to $WZ$ or dijets; and a $Z^\prime$ decaying to charged dileptons or dijets. We will first illustrate how to think about a few cases in terms of branching ratios, and then translate into making comparisons based on $\zeta$.  Subsequent examples will be explored in terms of $\zeta$ alone, because it is more versatile. We will start with cases where only one set of initial $(i,j)$ and final $(x,y)$ states is relevant and will close with an example of the more general case where multiple production or decay modes may contribute. 

\subsection{Fermiophobic $W^\prime$: $W_L^{\pm} Z_L \to W^{\prime} \to W_L^{\pm} Z_L$}
\label{subsec:fermiophobic-W-prime}

Our first example will be a charged spin-one color-neutral vector resonance -- a technirho or a $W^\prime$ -- that is primarily produced by vector boson fusion and primarily decays to $W_L^\pm Z_L$.  Because the initial and final states are identical (but the two incoming partons differ from one another), the signal cross section (in this simplified model) is determined entirely by $BR(R \to W_L Z_L)$, which cannot exceed 1.   

It is interesting to inquire whether such a resonance could have been responsible for the diboson excesses reported in the summer 2015 data from ATLAS and CMS (see, e.g., refs 1-14 of \cite{Brehmer:2015dan}). In the analysis reported in \cite{Aad:2015owa} based on hadronically-decaying dibosons, the most significant discrepancy in the $WZ$ channel from the background-only hypothesis occurs at an invariant mass of order 2 TeV; the local significance is $3.4\sigma$ and the global significance including the look-elsewhere effect in all three channels ($WZ, WW, ZZ$) is $2.5\sigma$.

\begin{figure}[btp]
	\centering
	\includegraphics[width=0.49\textwidth]{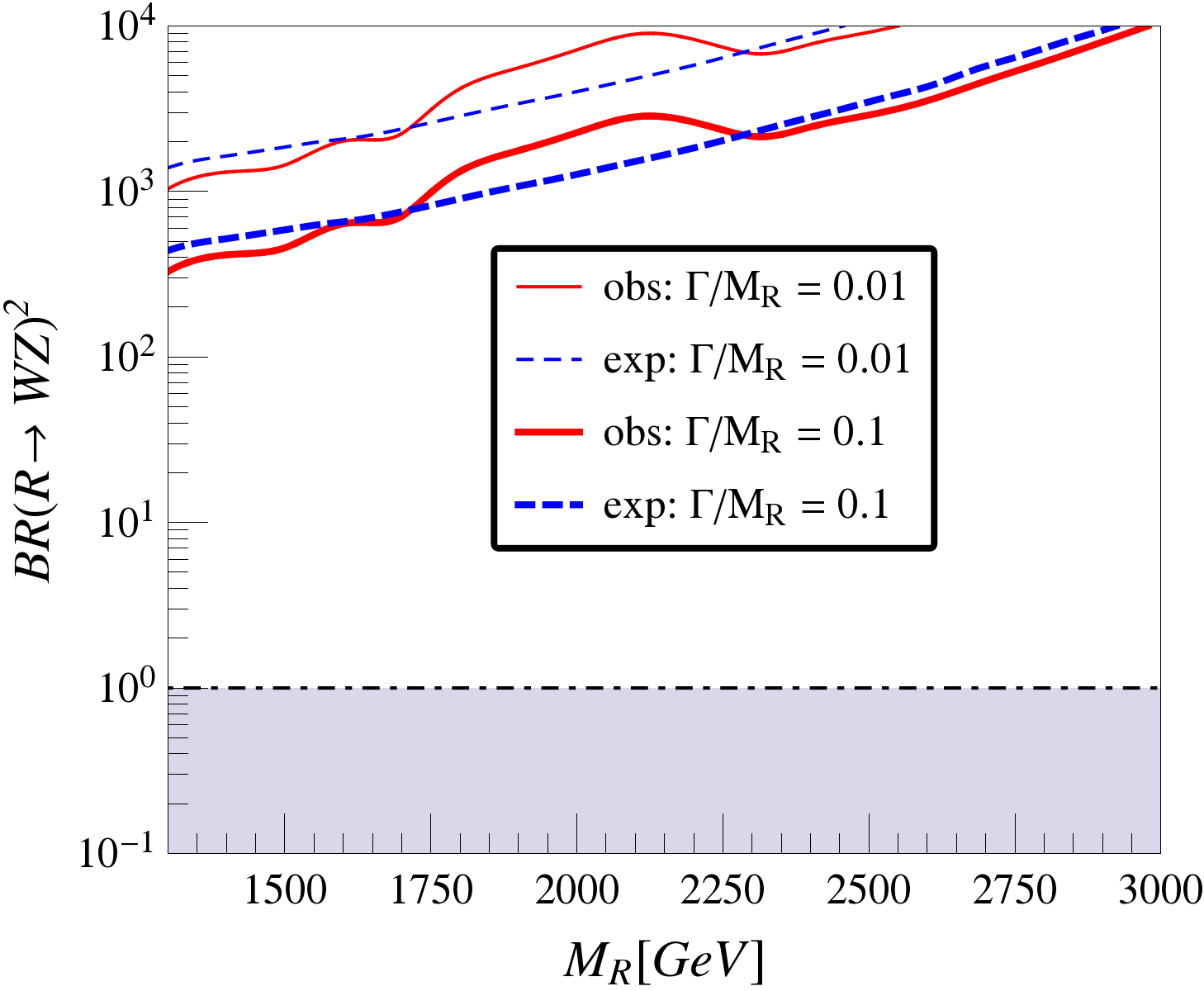}
	\includegraphics[width=0.49\textwidth]{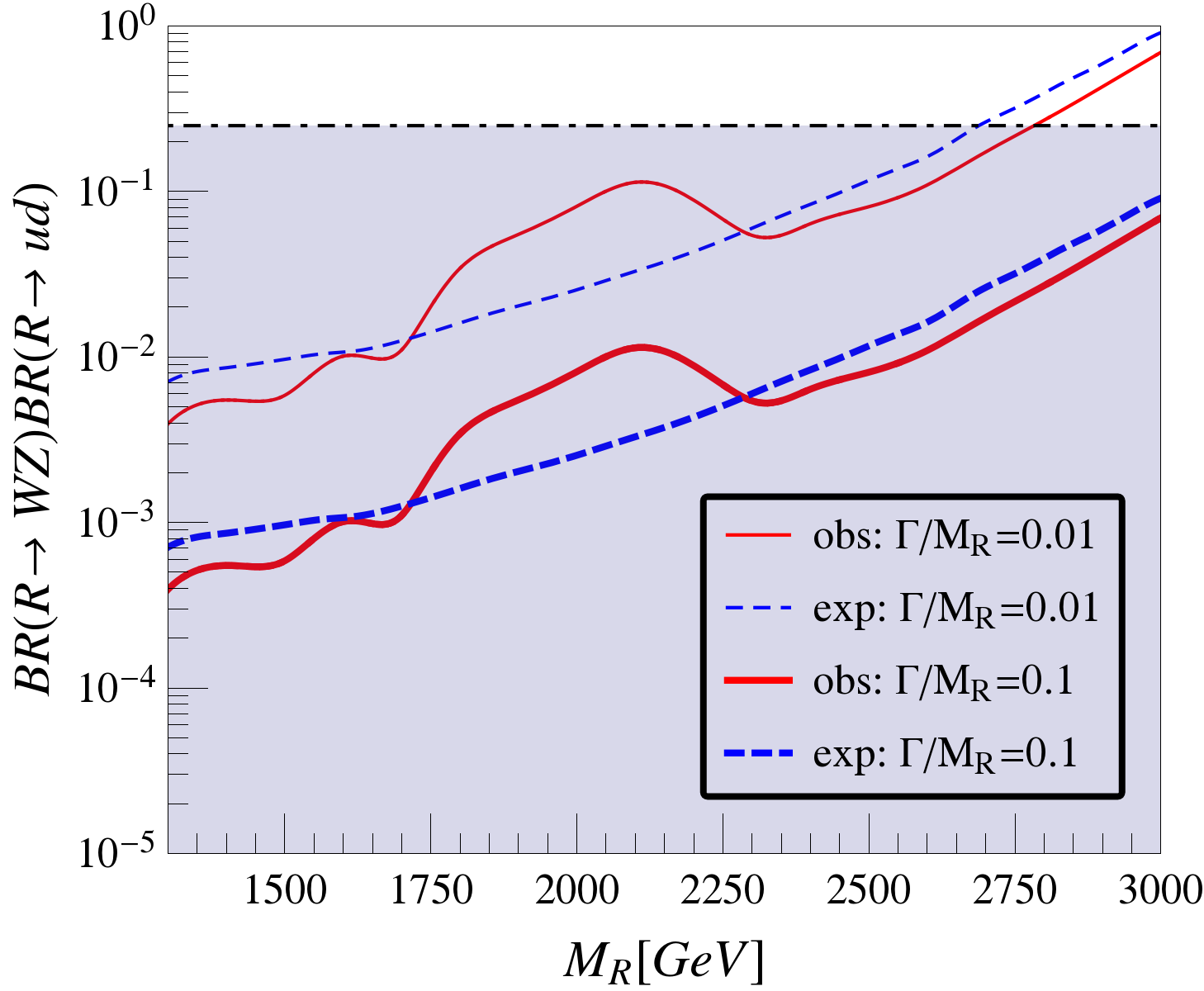}
	\caption{\small \baselineskip=3pt {\textbf{Left: }}The experimental ATLAS \cite{Aad:2015owa} upper limits (solid red curves) and expected limits (dashed blue curves) yield these upper bounds on the branching ratio product $[BR(R\to W_LZ_L)]^2$ assuming production of an s-channel resonance $R$ via vector boson fusion alone; results are shown for two values of $\Gamma/m_R$. As discussed in the text, since the apparent excess lies well outside the allowed (shaded) region, this scenario is disfavored. {\textbf{Right: }}The experimental ATLAS \cite{Aad:2015owa} upper limits (solid red curves) and expected limits (dashed blue curves)  yield these upper bounds on the branching ratio product $[BR(R\to u\bar{d} + d\bar{u})][BR(R\to W_LZ_L)]$ assuming production of an s-channel resonance $R$ via $u\bar{d} + d\bar{u}$ annihilation alone, shown for two values of $\Gamma/m_R$. As discussed in the text, since the apparent excess lies well within the allowed (shaded) region, this scenario was viable. }
	\label{fig:simplified-vbf2-ud2}
\end{figure}

In the left pane of ~\ref{fig:simplified-vbf2-ud2}, we have applied Eq. \ref{eq:simplebound-lower} to the observed and expected experimental upper limits \cite{Aad:2015owa} on the production cross-section for a resonance produced by $WZ$ fusion and decaying back to the same state.\footnote{We estimate the $WZ$ parton luminosities using the Effective W approximation~\cite{Dawson:1984gc,Dawson:1984gx,Kane:1984bb}; see also discussion in \cite{Chivukula:2016hvp}.} As this requires one to assume a specific value for the resonance's width/mass ratio, we show the results for $\Gamma_R / M_R = 1\%,\ 10\%$.   In the region around a resonance mass of 2 TeV, the observed upper bound is weaker than expected, indicating that an excess may be present.  From the plot, it is clear that the squared branching ratio $[BR(R \to WZ)]^2$ required to produce the excess production rate would be of order a few hundred to a few thousand. This greatly exceeds the maximum possible value of 1; allowed values of the squared branching ratio fall in the shaded region towards the bottom of the pane. Therefore, longitudinal vector boson fusion cannot be the dominant production mode for any $W^{\prime}$ resonance responsible for the observed possible diboson excess.  

\begin{figure}[btp]
	\centering
	\includegraphics[width=0.49\textwidth]{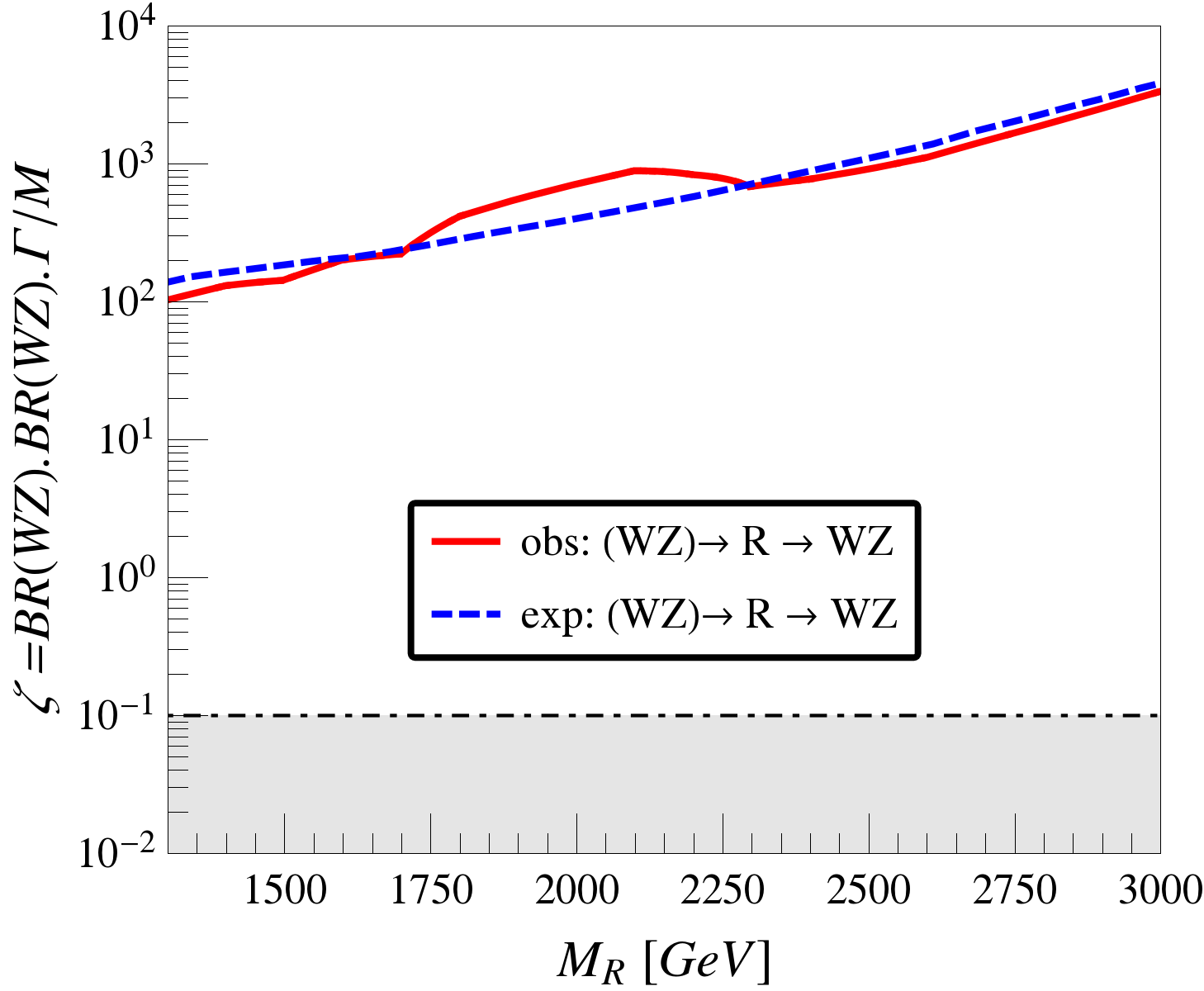}
	\includegraphics[width=0.49\textwidth]{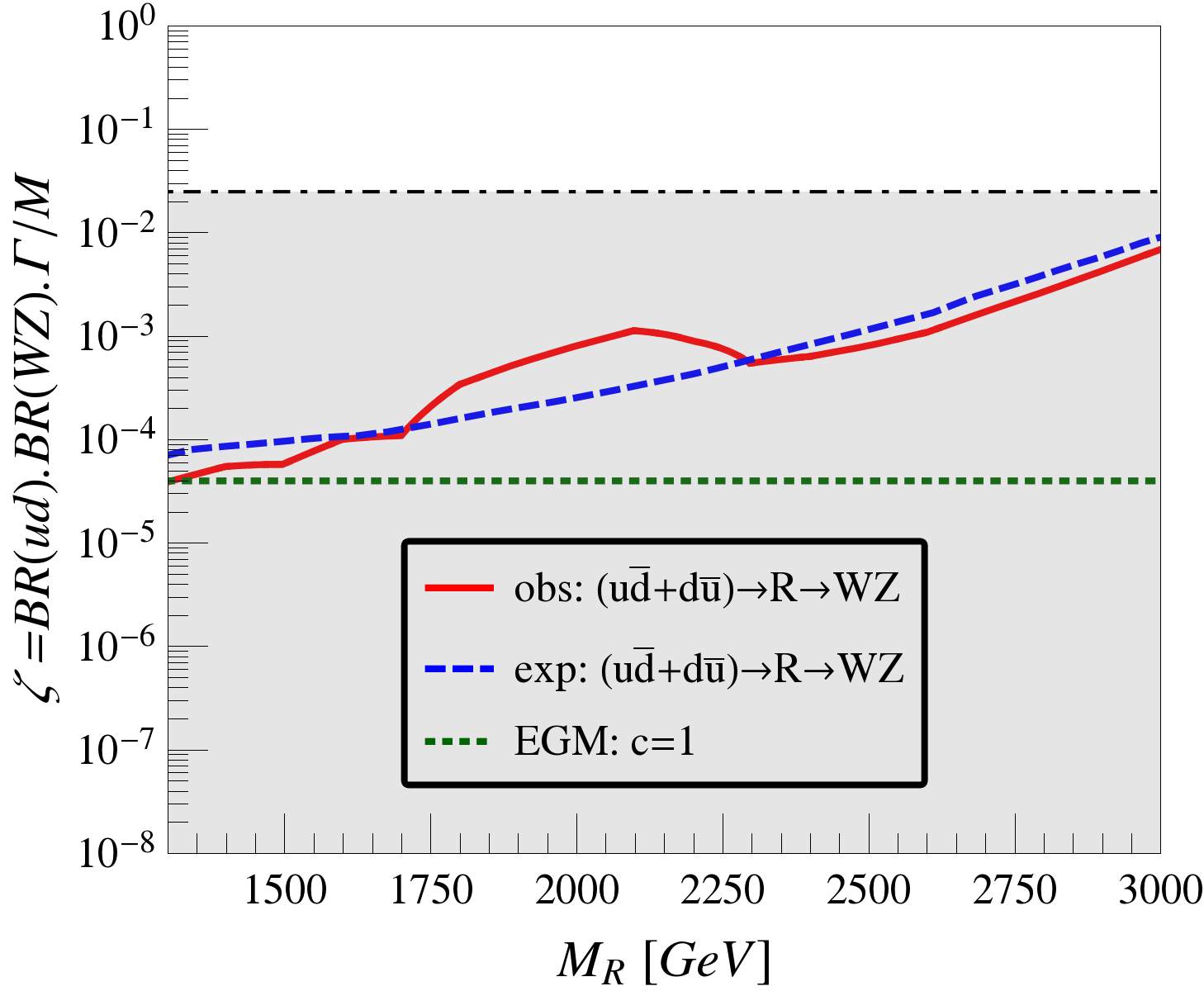}
	\caption{\small \baselineskip=3pt {\textbf{Left: }} The experimental ATLAS \cite{Aad:2015owa} upper limits and expected limits on the production cross-section for $WZ \to R \to WZ$ yield these upper bounds on $\zeta$ assuming production of the s-channel resonance $R$ via vector boson fusion alone. Since the potential excess (the difference between the curves) near 2 TeV corresponds to $\zeta \sim 100$, it lies well outside the allowed (shaded) region. {\textbf{Right: }}The experimental ATLAS \cite{Aad:2015owa} upper limits and expected limits on the production cross-section for $u\bar{d} + d\bar{u}) \to R \to WZ$ yield these upper bounds on $\zeta$ assuming production of the s-channel resonance $R$ via $u\bar{d} + d\bar{u}$ alone. Since the apparent excess corresponds to $\zeta \sim 10^{-4}$, it lies well within the allowed (shaded) region. The extended gauge model (EGM)~\cite{Altarelli:1989ff} with the coupling factor $c$ set to 1 predicts a value of eta well below what would be required to explain the apparent excess.}
	\label{fig:simplified-eta-vbf2-ud2}
\end{figure}

The same comparison is made in the left pane of Figure~\ref{fig:simplified-eta-vbf2-ud2} using the variable $\zeta$ on the vertical axis.  Using $\zeta$ removes the need to show separate curves for different values of $\Gamma/M$.  We see that the value of $\zeta$ corresponding to the possible excess production around 2 TeV would be $\zeta \sim 100$; this is far above the maximum value of 0.1 that forms the upper boundary of the shaded allowed region in the plot; hence, a fermiophobic resonance would not be a viable candidate for producing such an excess.

\subsection{Non-fermiophobic $W^\prime$ decaying to dibosons:  $u d \to W^{\prime} \to W^{\pm} Z$}
\label{subsec:nonfermiophoboc-dibosons}

Next, consider a charged spin-one color-neutral vector resonance -- a technirho or a $W^\prime$ -- that couples to both quark/anti-quark pairs and to vector boson pairs.  Since the parton luminosities for $(c,s,W^{\pm},Z)$ are small, those initial states can be neglected.
Thus, we have a resonance produced via $q\bar{q}$ (in this case primarily $u\bar{d}$ or $d \bar{u}$) annihilation and capable of decaying to vector boson pairs.  Since the 2015 data \cite{Aad:2015owa} showed a potential excess in diboson pairs but not one in dijets, we restrict ourselves to the situation in which the W' couples to $q\bar{q'}$ far more weakly than to $WZ$, so that dijet decays will not be significant.  Therefore, the $W'$ effectively has one production mode and a different single decay mode.

In this case, the signal cross section is determined entirely by $BR(R\to q\bar{q}) \cdot BR(R\to WZ)$, which is bounded from above by $1/4$, since the two incoming partons differ from one another.

In the right pane of Figure~\ref{fig:simplified-vbf2-ud2}, we have applied Eq.~\ref{eq:simplebound-lower} to the observed and expected experimental ATLAS upper limits on the production cross-section \cite{Aad:2015owa} to obtain an upper bound on the product of branching ratios of the resonance into the $ud$ initial state and  $W^\pm Z$ final state. As doing so requires one to assume a value for the resonance's width/mass ratio, we show the results for for $\Gamma_R / M_R = 1\%,\ 10\%$.  In contrast to the results for the fermiophobic $W'$ from the left pane of this figure, here we see that a $W^{\prime}$ produced via $q\bar{q}$ annhilation can be consistent with the observed excesses so long as the corresponding product of the branching ratios to $WZ$ and $q\bar{q}$ lies within the shaded region. 

The same comparison is made in the left pane of Figure~\ref{fig:simplified-eta-vbf2-ud2} using the variable $\zeta$ on the vertical axis.  This removes the need to show separate curves for different values of $\Gamma/M$.  Now we see that the value of $\zeta$ corresponding to the possible excess production around 2 TeV would be $\zeta \sim 10^{-4}$; this is well below the maximum value of 0.025 that forms the upper boundary of the shaded allowed region in the plot, leaving a non-fermiophobic $W^\prime$ as a viable possibility.  The $W^\prime$ boson of the extended gauge model (EGM)~\cite{Altarelli:1989ff} with the coupling factor $c$ set to 1 predicts a value of $\zeta$ well below what would be required to explain the apparent excess.

\subsection{Photophillic Resonance: $\gamma \gamma \to R \to \gamma \gamma$}
\label{subsec:gammas}

Let us move on to resonances that may be relevant to the hints of a new diphoton resonance at 750 GeV reported in winter 2015  \cite{ATLAS-Diphoton,Moriond-ATLAS,ATLAS-CONF-2016-018,CMS:2015cwa,CMS:2015dxe,Moriond-CMS,CMS-PAS-EXO-16-018}.  First, we consider a new state (either spin-0 or spin-2) produced by photon fusion and decaying only to diphotons.  Conceptually, this case resembles the fermiophobic $W'$ in that the unique initial and final states are identical; note, however, that the two initial state partons are identical, so that the upper limit on $(1+\delta_{ij})[BR(R\to \gamma\gamma)]^2$ is 2 rather than 1. This example would be of phenomenological interest if a new resonance were seen only in a diphoton decay channel.

\begin{figure}[htbp]
	\centering
	\includegraphics[width=0.49\textwidth]{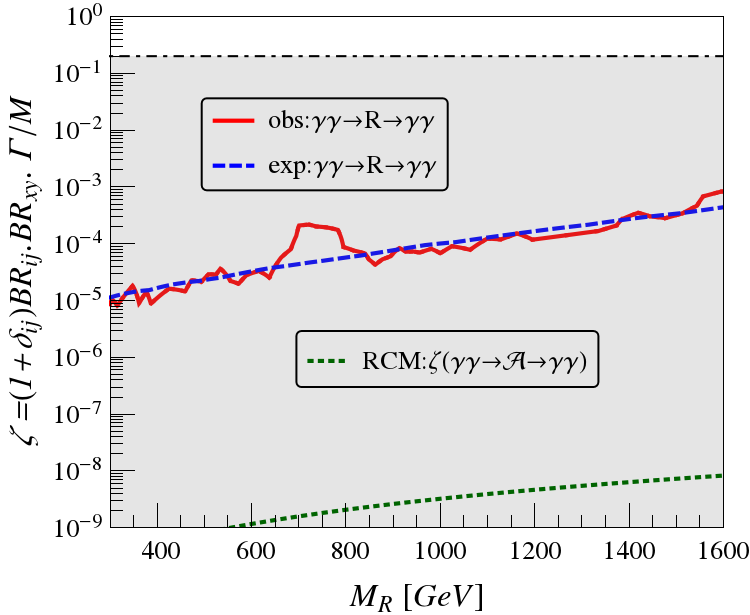}
		\includegraphics[width=0.49\textwidth]{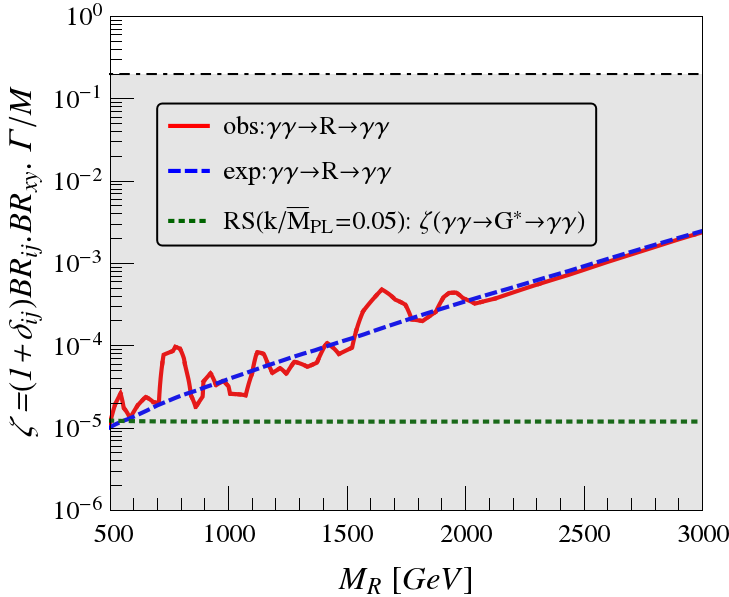}
	\caption{\small \baselineskip=3pt
Experimental observed (solid red) and expected (dashed blue) upper bounds~\cite{Aaboud:2016tru} on $\zeta$ for production of an s-channel  resonance $R$ via photon fusion and subsequent decay to diphotons. Since the contours lie in the (shaded) region where $\zeta$ is below the maximum value for this process, photon fusion alone may be the dominant production mode of such a narrow resonance. \textbf{		Left:} Spin-0 resonance. The green-dotted curve indicates the predicted value of $\zeta$ for the Renormalizable Coloron Model \cite{Chivukula:2016zbe}; it lies several orders of magnitude below the value required to account for the apparent excess events at 750 GeV.  
		\textbf{		Right:} Spin-2 resonance. The green-dotted curve indicates the predicted value of $\zeta$ for the RS Graviton model \cite{Bijnens:2001gh} with the parameter value as indicated; it lies about a factor of five below the value required to account for the apparent excess events at 750 GeV.
	}
	\label{fig:simplified-gaga1}
\end{figure}

The $\gamma \gamma$ luminosities are produced using the {\tt CT14} photon pdfs~\cite{Schmidt:2015zda}. Results are reported in  Figure~\ref{fig:simplified-gaga1}.  The solid curve shows the observed upper bound, while the expected upper bound is denoted by the dashed curve. We will take the difference between the observed and expected upper limits as an indication of the value of $\zeta$ required to produce the excess tentatively seen at a mass of 750 GeV; that points to a value of $\zeta$ of order $10^{-4}$. 

For comparison, the predicted value of $\zeta$ as a function of resonance mass in the Renormalizable Coloron Model (RCM)is shown (dotted green curve), assuming  that the pseudoscalar state is produced via diphoton fusion and decays back to diphotons.\footnote{The values of the parameters of the model are the same as in ref.~\cite{Chivukula:2016zbe}. The number of generations of singlet and doublet vector like quarks are chosen to be $n_q=3$ and $N_Q=3$.} The RCM was proposed in \cite{Bai:2010dj} and has also been studied extensively in \cite{Chivukula:1996yr,Hill:1993hs,Dicus:1994sw,Chivukula:2013xka,Chivukula:2014rka,Chivukula:2015kua,Chivukula:2016zbe}.   Unfortunately, we can see from the figure that the RCM would provide an $\zeta$ value five orders of magnitude too small.  A pseudoscalar produced by photon fusion in the RCM cannot account for the apparent excess.

 The right pane shows a comparison with the theoretically predicted value of $\zeta$ in the RS Graviton model \cite{Randall:1999ee, Bijnens:2001gh} for a spin-2 graviton produced by photon fusion and having $(k/\bar{M}_{Pl} = 0.05$ (dotted green curve). Since the prediction lies a factor of five below the value required to account for the apparent excess events at 750 GeV, it is not obvious that the model, with this choice of parameters, could provide an explanation; larger values of $k/\bar{M}_{Pl}$ may potentially accommodate the excess. 

\subsection{Boson-Phillic Resonance: $g g \to R \to \gamma \gamma$}
\label{subsec: bosophillic}

Alternatively, we may consider a (spin-0 or spin-2) resonance that can be produced via gluon fusion and still decays to photons. Conceptually, this resembles the non-fermiophobic $W'$ in that the unique initial and final states differ from one another.  Since the two incoming partons are identical, the upper limit on $(1+\delta_{ij})BR(R\to gg) BR(R\to\gamma\gamma)$ is $1/2$.  Again, this case is of interest if a diphoton resonance is seen without a corresponding dijet signal; the resonance's branching fraction to dijets must be small enough to avoid a dijet signal yet still large enough that gluon fusion is the dominant production mode. 

\begin{figure}[htbp]
	\centering
	\includegraphics[width=0.49\textwidth]{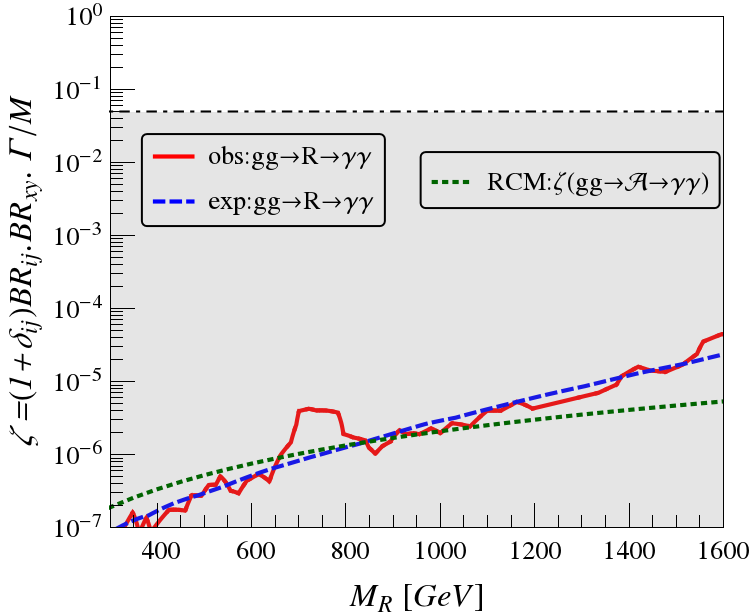}
		\includegraphics[width=0.49\textwidth]{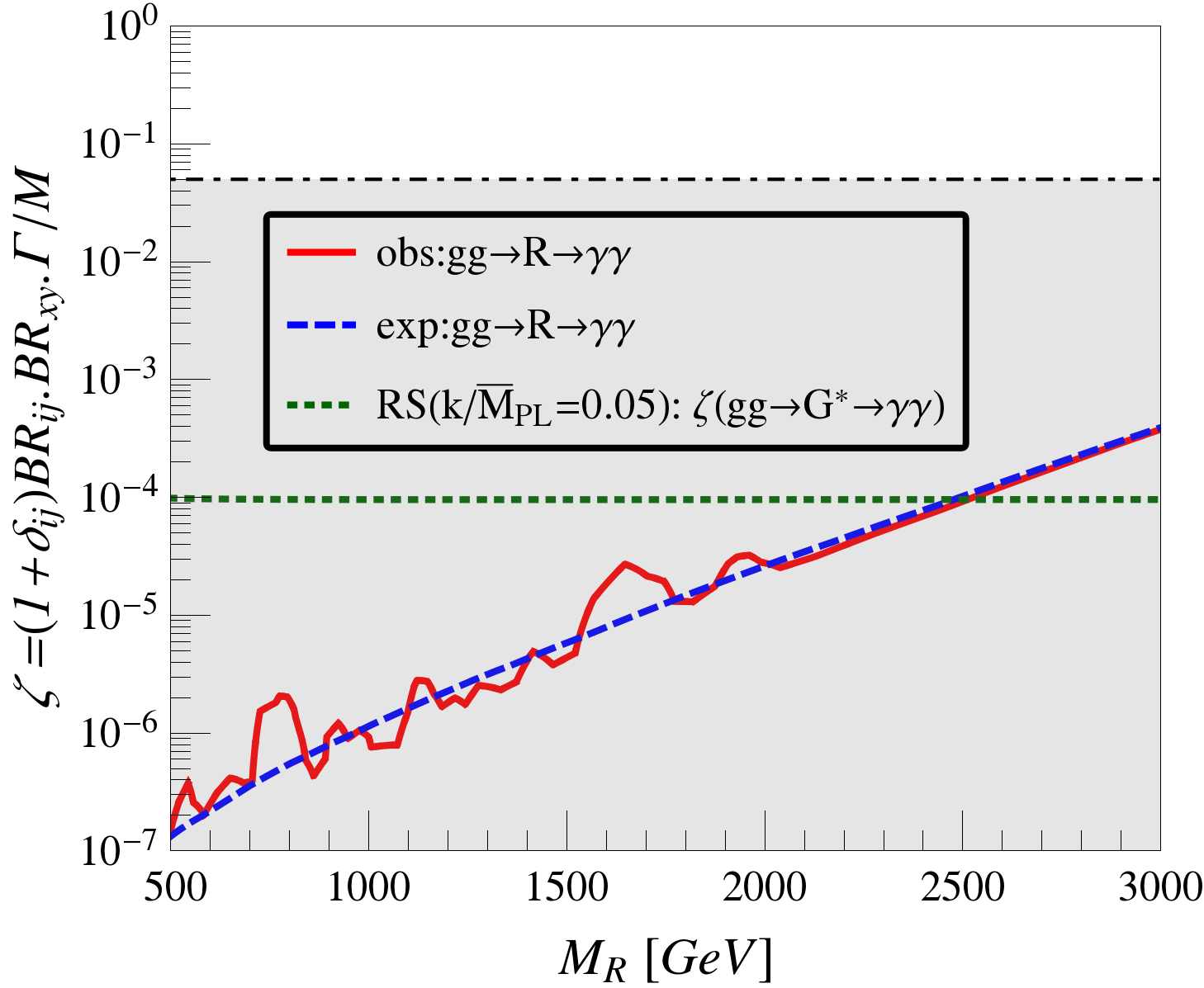}
	\caption{\small \baselineskip=3pt
		Experimental observed (solid red) and expected (dashed blue) upper bounds~\cite{Aaboud:2016tru} on $\zeta$ for production of an s-channel  resonance $R$ via gluon fusion and subsequent decay to diphotons. Since the contours lie in the (shaded) region where $\zeta$ is below the maximum value for this process, gluon fusion alone may be the dominant production mode of such a narrow resonance. \textbf{		Left:} Spin-0 resonance. The green-dotted curve indicates the predicted value of $\zeta$ for the Renormalizable Coloron Model \cite{Chivukula:2016zbe}; it crosses through the window between the observed and expected upper bounds on eta at a resonance mass of order 750 GeV, indicating that the RCM could explain the apparent excess of diphoton events.  
		\textbf{		Right:} Spin-2 resonance. The green-dotted curve indicates the predicted value of $\zeta$ for the RS Graviton model \cite{Bijnens:2001gh} with the parameter value as indicated; it is excluded for resonance masses below about 2.5 TeV, setting a lower bound on the graviton mass.  
	}
	\label{fig:simplified-gg1}
\end{figure}

Results are reported in  Figure~\ref{fig:simplified-gg1}.  The solid (red) curve shows the observed upper bound, while the expected upper bound is denoted by the dashed (blue) curve. Note that the upper bound on $\zeta$ as a function of resonance mass is far more stringent for a resonance produced by gluon fusion (Figure~\ref{fig:simplified-gg1}) than for one produced by photon fusion (Figure~\ref{fig:simplified-gaga1})), because the gluons' parton luminiosity is so much larger.  Similarly, because the PDFs of the gluon and photon have different energy dependences, the slopes of the upper bound curves are also slightly different from one another.
 
 For comparison, in the left pane the predicted value of $\zeta$ as a function of resonance mass in the Renormalizable Coloron Model (RCM)  is shown (dotted green curve), assuming that the pseudoscalar state characteristic of that model is produced via gluon fusion and decays back to diphotons.\footnote{For the RCM, we choose the same parameter values as in ref.~\cite{Chivukula:2016zbe}, and set the number of generation of doublets and singlet to be three each $(N_Q = n_q =3)$.}   If we take the difference between the observed and expected upper limits as an indication of the value of $\zeta$ required to produce the excess tentatively seen at a mass of 750 GeV, that points to a value of $\zeta$ of order $10^{-6}$, which is in line with the RCM prediction. Thus, as discusssed in Ref.~\cite{Chivukula:2016zbe}, the RCM is a viable candidate model to explain such a diphoton excess.   
 
 In the right pane, comparison with the theoretically predicted value of $\zeta$ in the RS Graviton model with $(k/\bar{M}_{Pl} = 0.05$ is shown (dotted green curve).  This illustrates that the RS graviton predicts a value of $\zeta$ that is excluded for resonance masses below about 2.5 TeV, setting a lower bound on the graviton mass. It is therefore not a good candidate to explain a diphoton excess at 750 GeV (the predicted value of $\zeta$ lies several orders of magnitude above the upper bound at that mass). 

\subsection{$Z^{\prime}$ boson: $u\bar{u}, d\bar{d} \to Z^{\prime} \to jj, b\bar{b}, \ell^+\ell^-$}
\label{subsec: Zprime}

Production of a $Z^{\prime}$ boson exemplifies th general resonance scenario, since many proposed $Z^{\prime}$ bosons have significant couplings to both up and down quarks, giving $u\bar{u}$ and $d\bar{d}$ annihilation as separate production channels with distinct parton luminosities. As noted in \cite{Chivukula:2016hvp}, the upper bound on $\zeta$ depends on the combination of production and branching modes that are relevant for the particular search. 

\begin{figure}[htbp]
	\centering
	\includegraphics[width=0.48\textwidth]{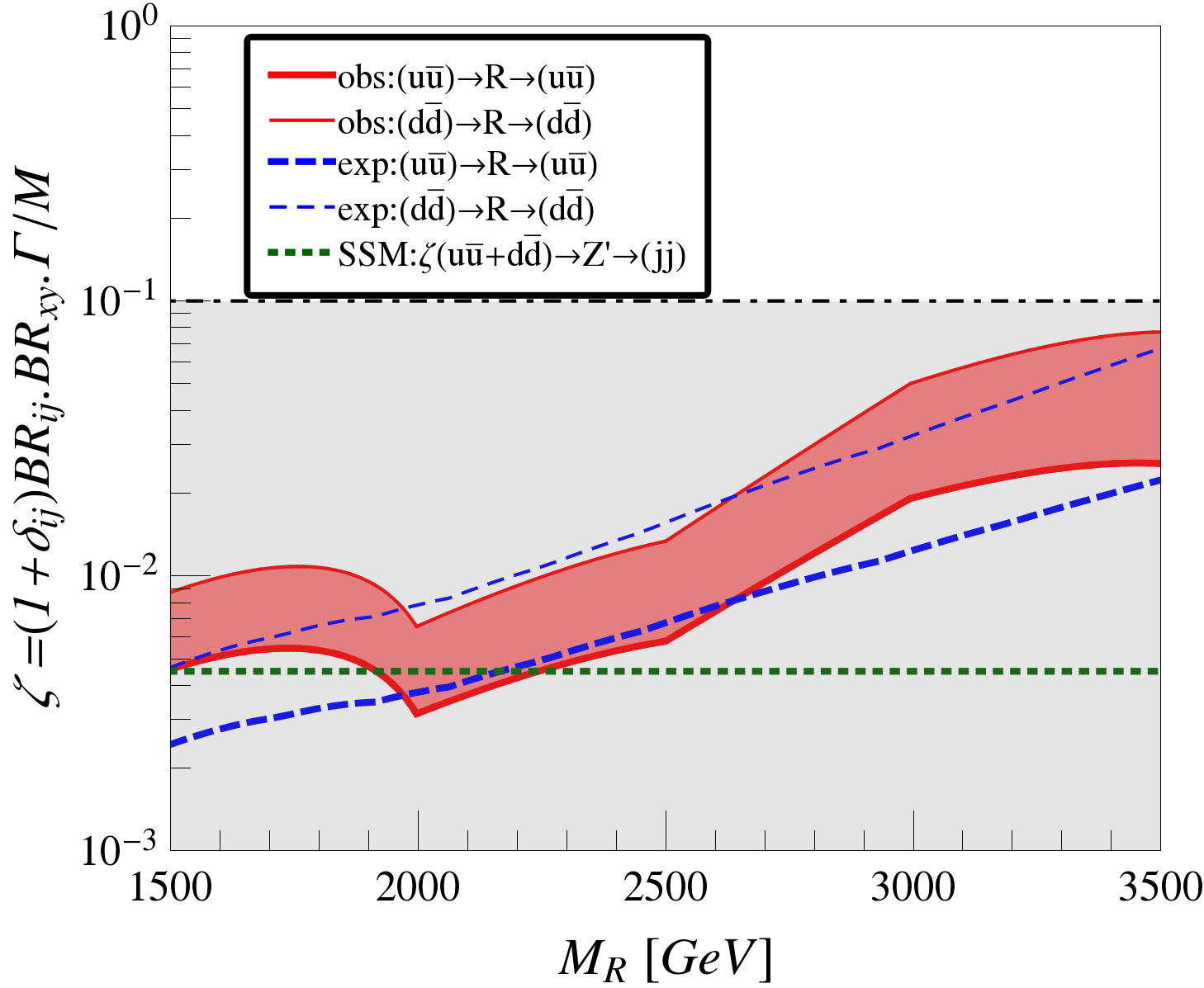}
	\includegraphics[width=0.48\textwidth]{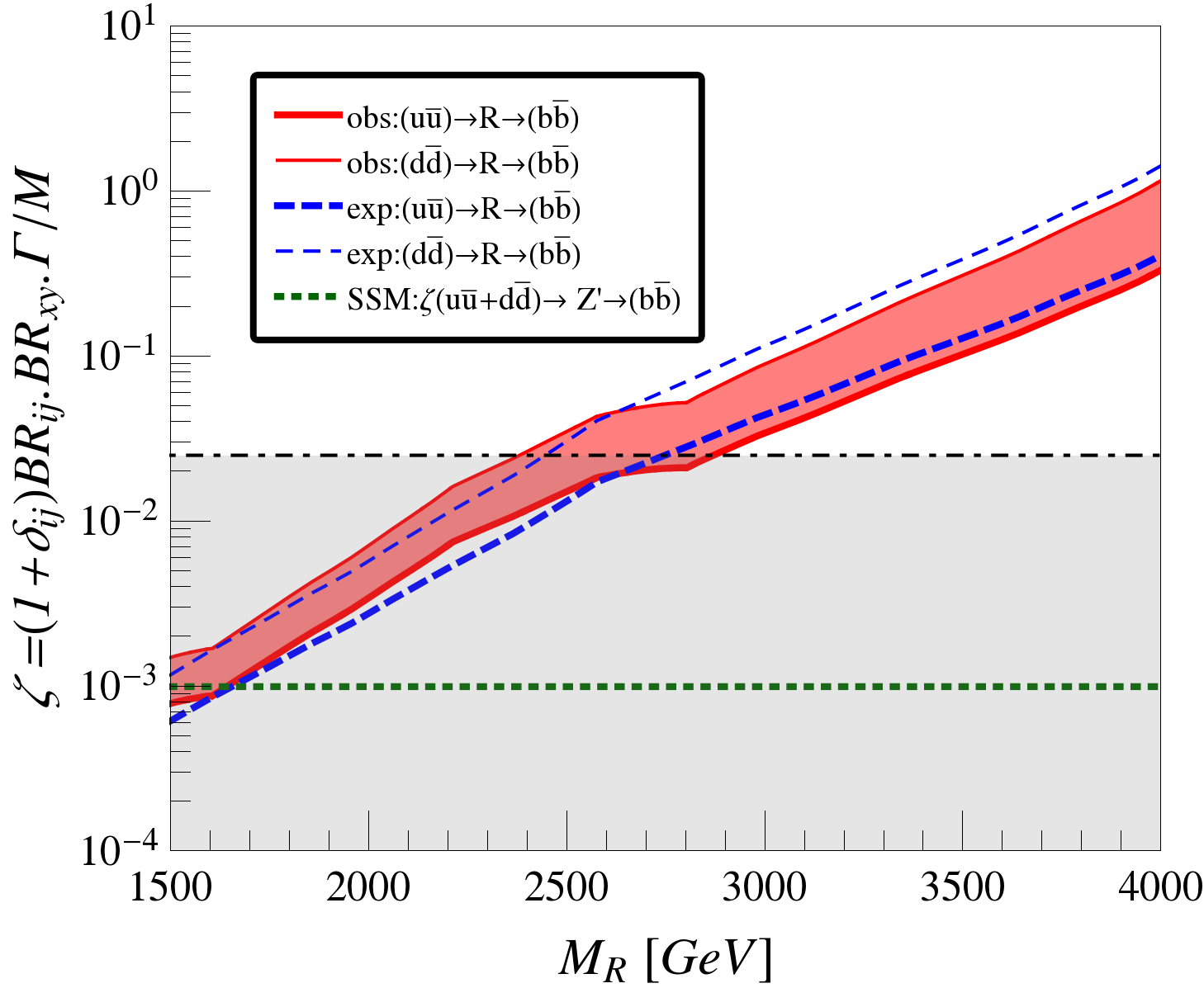}\vspace{0.5cm}
		\includegraphics[width=0.48\textwidth]{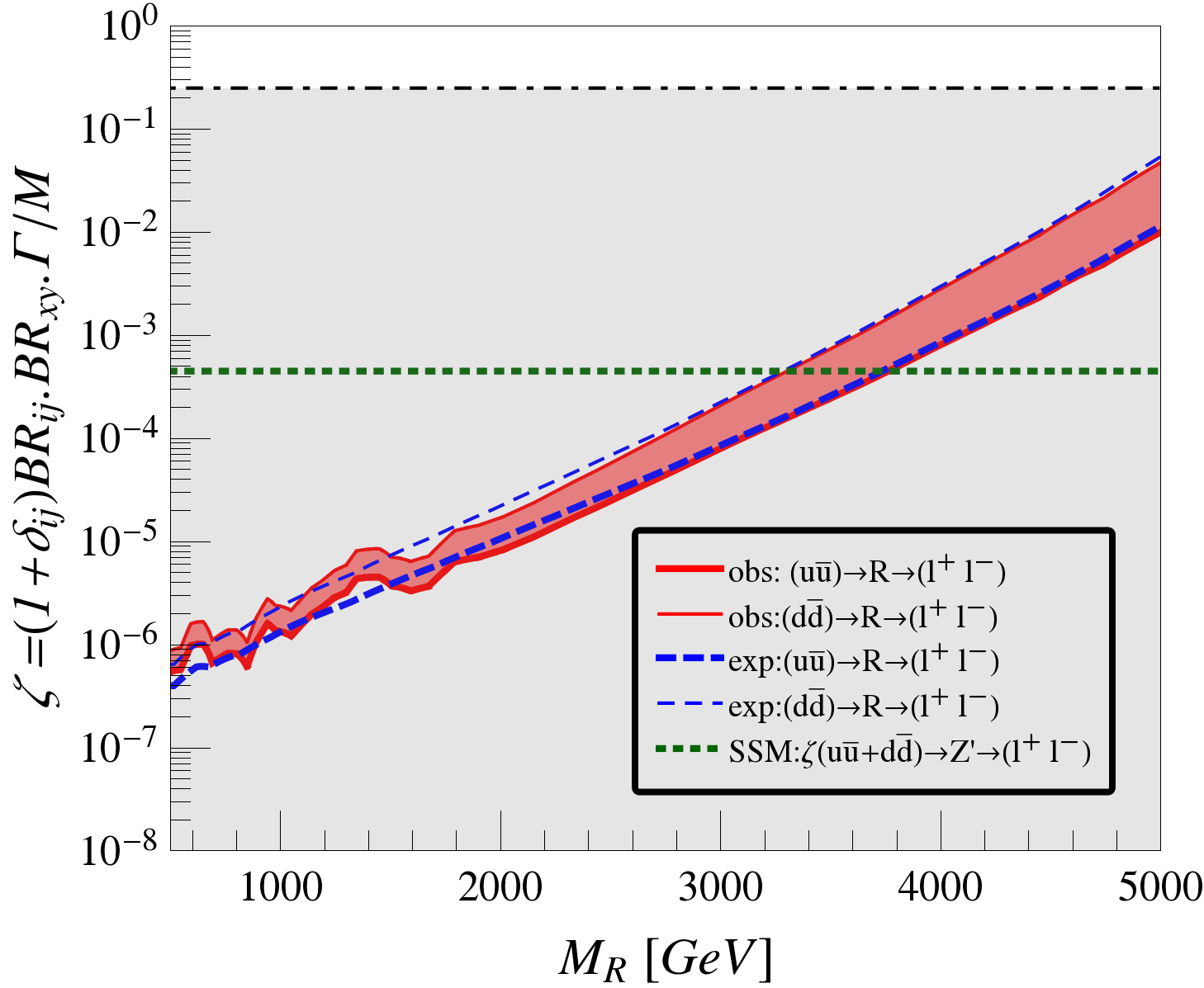}
	\caption{\small \baselineskip=3pt  In these plots in the $\zeta$ vs. resonance mass plane, the shaded region corresponds to values of $\zeta$ consistent with upper bounds on branching ratios as described in \cite{Chivukula:2016hvp} (1 for the upper left pane and 1/4 for the other panes) and $\Gamma/M \leq 10\%$. The observed (red solid) and expected (blue dashed) upper bounds on $\zeta$ as a function of resonance mass are shown in each pane.  The thick solid red and blue dashed curves correspond to the situation in which the $Z^\prime$ couples only to up-flavor (down-flavor) quarks.  The shaded band between the two red curves represents the range of variation of the upper bound on $\zeta$ as the $Z^\prime$ ranges between the coupling extremes represented by the two red curves.
		\textbf{	Upper	Left:}
				Upper bounds on $\zeta$ for a leptophobic $Z^{\prime}$ produced via light quark/anti-quark annihilation and decaying to dijets (red solid curve) compared with expected upper limit (blue dashes) and the size of $\zeta$ provided by a Sequential Standard Model $Z^\prime$ boson (green dots). If a significant excess were deemed present at masses below 2 TeV, the contribution of this $Z^\prime$ boson would be consistent with it.
		\textbf{	Upper	Right: }Upper bounds on $\zeta$ for a leptophobic $Z^{\prime}$ produced via light quark/anti-quark annihilation and decaying to third generation quarks, compared with expected upper limit (blue dashes) and SSM $Z^\prime$ (green dots). 	\textbf{		Lower: }Upper bounds on $\zeta$ for a  $Z^{\prime}$ produced via light quark/anti-quark annihilation and decaying to dileptons, compared with expected upper limit (blue dashes) and SSM $Z^\prime$ (green dots).
	}
	\label{fig:simplified-Zp}
\end{figure}

To facilitate the comparison with models, we show $\zeta$ on the vertical axis of the plots showing our results.  Since we are studying narrow resonances, with $\Gamma_R / M_R \leq 10\%$, the upper bound on $\zeta$ in each of the three cases discussed above would be one tenth the bound on the product of branching ratios. The shaded region in each plot of Figure~\ref{fig:simplified-Zp} corresponds to the region obeying that bound in the $\zeta$ vs. resonance mass plane. The observed (red solid) and expected (blue dashed) upper bounds on $\zeta$ as a function of resonance mass are shown in each pane.  The thick (thin) solid red and blue dashed curves correspond to the situation in which the $Z^\prime$ couples only to up-flavor (down-flavor) quarks.  The shaded band between the two red curves represents the range of variation of the observed upper bound on $\zeta$ as the $Z^\prime$ ranges between the coupling extremes represented by the two red curves.  This covers the full range of possibilities for $Z^\prime$ bosons coupling to first-generation quarks.  

The upper left pane of Figure~\ref{fig:simplified-Zp} shows the observed upper bounds on $\zeta$ for a leptophobic $Z^{\prime}$ produced via light quark/anti-quark annihilation and decaying to dijets (red solid) alongside the expected upper limit (blue dashes) \cite{ATLAS:2015nsi}.  The value of  
 $\zeta$ for a Sequential Standard Model (SSM) $Z^\prime$ boson ~\cite{Langacker:2008yv} (green dots) is shown for comparison. 
If one suspected that the difference between the observed and expected upper limit near 1.75 TeV, for instance, corresponded to an excess of events stemming from the presence of a resonance, then the SSM $Z^\prime$ would provide a value of $\zeta$ consistent with that required of the resonance.  However, if one made a similar comparison around 3 TeV, it would be clear that the SSM $Z^\prime$ had too small an $\zeta$ value to be the source of such a postulated excess.
 
The upper right pane of Figure~\ref{fig:simplified-Zp} shows the upper bounds on $\zeta$ for a leptophobic $Z^{\prime}$ produced via light quark/anti-quark annihilation and decaying to b-quarks (red solid) alongside the expected upper limit (blue dashes) \cite{Aaboud:2016nbq} and the value of $\zeta$ for the SSM $Z^\prime$ (green dots). Here, the value of $\zeta$ from the model lies well below the current upper limits for resonance masses above about 2 TeV.  Note also that the values of $\zeta$ probed by the data for resonance masses above about 3 TeV lie outside the allowed (shaded) region -- and, hence, if an excess of $b\bar{b}$ events with an invariant mass had been observed in this region, it could not have arisen from a model of this type. We find that the SSM $Z^\prime$ is not bounded by experiments  for the mass range shown in the figure. This is consistent with the results in  Ref.~\cite{Aaboud:2016nbq}.

The lower pane of Figure~\ref{fig:simplified-Zp} shows the upper bounds on $\zeta$ for  a $Z^{\prime}$ produced via light quark/anti-quark annihilation and decaying to dileptons (red solid) alongside the expected upper limit (blue dashes) \cite{ATLAS-CONF-2015-070} and the value of $\zeta$ for the SSM $Z^\prime$ (green dots).  Here, the value of $\zeta$ provided by the model is excluded by the data for masses below about 3.5 TeV; this provides a lower bound on the SSM $Z^\prime$ boson mass. A similar bound of about 3.5 TeV, is also obtained for the SSM $Z^\prime$ in Ref.~\cite{ATLAS-CONF-2015-070}.

\section{Discussion}
\label{sec:disc}

This talk has summarized a method first reported in \cite{Chivukula:2016hvp} for quickly determining whether a small excess observed in collider data could potentially be attributable to the production and decay of a single, relatively narrow, s-channel resonance belonging to a generic category, such as a leptophobic $Z^\prime$ boson or a fermiophobic $W^\prime$ boson. Using a simplifed model of the resonance allows us to convert an estimated signal cross section into general bounds on the product of the dominant branching ratios corresponding to production and decay. This quickly reveals whether a given class of models could possibly produce a signal of the required size at the LHC and circumvents the present need to make laboreous comparisons of many individual theories with the data.  Moreover, the dimensionless variable $\zeta$, which multiplies the product of branching ratios by the width-to-mass ratio of the resonance as defined in Eqn.~\ref{eq:gen-bran-rat}, does an even better job at producing compact and easily interpretable results.

In this work, we began by setting up the general framework for obtaining simplified limits and outlining how it applies for narrow resonances with different numbers of dominant production and decay modes.  We then analyzed applications of current experimental interest, including resonances decaying to dibosons, diphotons, dileptons, or dijets. As discussed in \cite{Chivukula:2016hvp}, further extensions of this work to cases that go beyond narrow resonances are now underway.

If the LHC experiments report their searches for BSM resonances in terms of the simplified limits variable $\zeta$, alongside the commonly used $\sigma \cdot BR$ now employed, this would make it far easier to discern what sorts of BSM physics might underly any observed deviations from SM predictions.

\section*{Acknowledgments}

 The work of. R.S.C., K.M., and  E.H.S. was supported by the National Science Foundation under Grants PHY-0854889 and PHY-1519045.  R.S.C. and E.H.S. also acknowledge the support of NSF Grant PHYS-1066293 and the hospitality of the Aspen Center for Physics during work on this paper. P.I. is supported by ``CUniverse" research promotion project by Chulalongkorn University (grant reference CUAASC).

\bibliography{SIMMONS_Elizabeth_CONF12}

\end{document}